\documentstyle[aps,preprint,epsf,amsmath]{revtex}

\begin{document}

\draft

\title{Further study of the Over-Barrier Model to compute charge exchange 
processes}

\author{Fabio Sattin \thanks{E-mail: sattin@igi.pd.cnr.it}}

\address{Consorzio RFX, Corso Stati Uniti 4, 35127 Padova, ITALY}

\maketitle

\abstract{
In this work we present an improvement over the Over Barrier Model 
(OBM) described in a recent paper [F. Sattin, Phys. Rev. 
A {\bf 62}, 042711 (2000)]. We show that: i) one of the two free parameters 
there introduced actually comes out consistently from the starting 
assumptions underlying the model; ii) the modified model thus 
obtained is as much accurate as the former one. Furthermore, we show 
that OBMs are able to accurately predict some recent results of state 
selective electron capture, at odds with what previously supposed.}
 

The electron capture process in collisions of slow 
ions with neutral atoms or other ions is of great importance  
in basic atomic physics, plasma physics and astrophysics. 
In principle, one could compute all the quantities of interest in 
such processes by writing
the time-dependent Schr\"odinger equation for the system 
and programming a computer to solve it. 
This task can be performed on present--days supercomputers for moderately 
complicated systems. 
Notwithstanding this, simple approximate models are still valuable:
(i) they allow getting analytical estimates, which are easy to adapt to 
particular cases; 
(ii) allow getting physical insight on the features of the problem by looking at
the analytical formulas; 
(iii) finally, they can be the only tools available when the complexity 
of the problem overcomes the capabilities of the computers. 
For this reason new models are being still developed 
\cite{niehaus,ostrovsky,jpb}. \\
The present author has presented in a recent paper \cite{pra} a study 
attempting to improve an already existing OBM \cite{ostrovsky} (this 
model will be hereafter referred to as I).  
The model there developed is able to predict cross sections for 
electron capture and ionization with appreciable accuracy for a large 
number of test cases. 
The key element was found to be the inclusion within the model of two 
free parameters, there labeled $\alpha$ and $f_{T}$. 
A large part of the paper \cite{pra} was devoted to show that, more than simple adjustable parameters,  
$\alpha$ and $f_{T}$ stand for some physical mechanism still not adequately 
included within the model. As such, one should expect they retain 
constant values from case to case, or vary according to some well defined 
relationship with the parameters of the problem at hand. 
Actually, it was found, by applying the model to a number of test 
cases, that a good agreement with experiment and/or other 
computations was obtained always with the same choice for both parameters
(in detail, $\alpha = 1$, $f_{T} = 2$). \\
In this paper we show that a 
correction to the capture probability, having the same meaning of parameter 
$f_T$, should appear naturally within the framework of the model I; that in 
the work \cite{pra} it was incorrectly overlooked and, as a 
consequence, we were forced to insert $f_{T}$ by hand in order of recovering 
accuracy of the results. 
 
Let us begin with a brief summary of model I; the reader is referred to \cite{pra} for a more complete 
discussion. 
We consider a standard scattering experiment between a target nucleus {\bf T}
and a projectile nucleus {\bf P} with only one active electron {\bf e}. 
We are considering hydrogenlike approximations for both the target and the 
projectile. Let {\bf r} 
be the electron vector relative to {\bf T} and {\bf R} the 
distance between {\bf T} and {\bf P}. In the spirit of classical OBMs, all particles
are considered as classical objects. \\ 
Let us consider the plane ${\cal P}$ containing all the three particles and use 
the projection of cylindrical polar coordinates $(\rho, z, \phi \equiv 0)$ to 
describe the position of the electron within this plane. We can 
assign the $z$ axis to the direction along the internuclear axis. \\ 
The total energy of the electron is (atomic units will be used unless 
otherwise stated):
\begin{equation}
\label{eq:uno}
E = {p^{2} \over 2 } + U = {p^{2} \over 2 }
-{ Z_{t} \over \sqrt{\rho^{2}+z^{2} } }- {Z \over \sqrt{\rho^{2}+(R-z)^{2}}} 
\quad .
\end{equation}
$Z$ and $Z_{t}$ are the effective charge of the projectile and  
of the target, respectively.  
From here on, we assign an effective charge $Z_t = 1$ to the target and
an effective quantum number $n$ to label the binding energy of the 
electron: $ E_{n} \equiv  1/(2 n^{2}) $. \\
When the projectile approaches the target nucleus, it also contribute 
to increase (in absolute value) the binding energy of the electron:
for distant encounters, we can approximate $E$ as 
\begin{equation}
\label{eq:due}
E(R) = - E_{n} - {Z \over R} \quad .
\end{equation}
On the plane ${\cal P}$ we can draw a section of the equipotential surface 
\begin{equation}
\label{eq:equip}
 U(z,\rho,R) = - E_n  - {Z \over R} \quad . 
\end{equation}
This represents the limit of the region classically allowed to the electron.
When $R \to \infty$ this region is divided into two 
disconnected circles centered on each of the two nuclei. Initial
conditions determine which of the two regions actually the electron lives in.
As $R$ diminishes there can be eventually an instant where the two regions
become connected. See fig. (1) of \cite{pra} for an example of this. \\
In the spirit of OBMs it is the opening of
the equipotential curve between {\bf P} and {\bf T}
which leads to a leakage of electrons from one nucleus to another, and
therefore to charge exchange.
It is easy to solve eq. (\ref{eq:equip}) for $R$ by imposing a vanishing width of the opening: 
\begin{equation}
\label{eq:rm}
R_{m} = { (1 + \sqrt{Z})^{2} - Z \over E_{n} } \quad .
\end{equation} 
In the region of the opening the potential $U$ has a saddle structure.
Charge loss occurs provided the electron is able to cross this potential 
barrier. 
Let $N_{\Omega}$ be the fraction of trajectories which lead to electron loss at the time $t$.
An approximate expression (valid for distant collisions) for $N_{\Omega}$ 
is given in \cite{ostrovsky}.
We simply quote that result:
\begin{equation}
\label{nomega}
N_{\Omega} \approx {1 \over 2} { \sqrt{Z} \over (\sqrt{Z} 
+ 1)^2} \left[  (\sqrt{Z} + 1)^2  - Z - E_n R \right] \quad .
\end{equation}
The leakage probability is related to $N_{\Omega}$ through
\begin{equation}
\label{eq:prob}
P_l =1 - \exp \left( - {f_{T} \over T} \int_{- t_m}^{+ t_m} N_{\Omega} dt \right) \quad .
\end{equation}
In this expression $dt/T$ is the fraction of electrons 
which cross any surface perpendicular to their motion (and enter the loss 
region) within time interval $dt$, with $T = 2 \pi n^{3}$
the unperturbed period of the electron motion along its orbit, 
and $f_{T}$ a corrective term which accounts for the perturbation.\\
In order to actually integrate Eq. (\ref{eq:prob}) we need to know the 
collision trajectory; for this an unperturbed straight line with $b$ impact parameter 
is assumed:
\begin{equation}
\label{eq:traiettoria}
R = \sqrt{b^{2} + (v t)^{2}} \quad .
\end{equation}
The extrema $ \pm t_m$ in the integral (\ref{eq:prob}) are the maximum values of 
$t$ at which charge loss can occur. They are related through Eq. 
(\ref{eq:traiettoria}) to the maximum 
distance at which capture may occur, $R_{m}$ (Eq. \ref{eq:rm}). 
This is the original estimate for $R_{m}$ as given in \cite{ostrovsky}. 
In \cite{pra} 
this estimate was questioned on the basis of the fact that it 
overestimated the maximum impact parameter available for charge 
exchange as computed by Classical Trajectory Monte Carlo (CTMC) calculations. 
As a consequence, the cross sections were overestimated, too. 
To remedy this, in \cite{pra} it was suggested to replace Eq. 
(\ref{eq:rm}) with
\begin{equation}
\label{eq:rcap2}
R'_{m} =  { (\alpha \sqrt{Z} + 1) \over E_{n} } \quad .
\end{equation}
With the choice $\alpha = 2$ we recover Eq. (\ref{eq:rm}), but it was 
found that better agreement with data was obtained for $\alpha = 1$.  
The value $\alpha = 1$ can be given also a physical meaning: it is 
easy to show (see for details ref. \cite{pra}) that, when substituted 
into Eq. (\ref{eq:rcap2}), it yields the maximum distance at which an 
electron can be captured provided that, prior to the capture, the 
electron trajectory is not perturbed in any way by the projectile, i.e.
the electron follows a trajectory with constant energy $ E = - E_{n}$, 
instead of $E$ given by  eq. (\ref{eq:due}).

We can write, after all this,
\begin{equation}
\begin{split}
\label{eq:effe}
&\int_{-t_m}^{t_{m}} N_{\Omega} dt = 2 F \left({ v t_{m} \over b }\right)  \\
F(u)  &= { \sqrt{Z}   \over  2 (\sqrt{Z} + 1)^2 }
\left[  \left( (\sqrt{Z} + 1)^{2} - Z \right)  { b \over v} u -  
\left({E_n b^{2} \over 2 v}\right) \left( u \sqrt{ 1 + u^{2}} + 
{\rm arcsinh}(u) \right) \right] \quad .
\end{split}
\end{equation}
The cross section can be finally obtained after integrating over the impact
parameter (this last integration must be done numerically):
\begin{equation}
\label{eq:sigma}
\sigma = 2 \pi \int b P_l(b) db \quad .
\end{equation}
The integration extends till the maximum $b$ allowed: $ b_{m} = R'_{m}$.

The key point we want to underline here is that the definition of the 
orbital period given above
is not consistent with basic hypotheses (\ref{eq:due}): it is based 
in fact on the relation for the periodic motion along the radial 
direction \cite{landau}:
\begin{equation}
\label{eq:landau}
T = 2 \int_0^{1/E}{dr \over p} = 
\sqrt{2} \int_0^{1/E} {dr \over \sqrt{ {1 \over r} - E}}  \quad .
\end{equation}
One recovers $T = 2\pi n^{3}$ by putting $E \equiv E_{n}$ in 
this equation. However, 
to be consistent with Eq. (\ref{eq:due}) one should assume that the orbital
period of the electron is changed, just like its binding 
energy, while the projectile is approaching. The expression Eq. (\ref{eq:due})
should thus be used in (\ref{eq:landau}).  
By doing so, one gets
\begin{equation}
\label{eq:periodprime}
T'  = 2 \pi \left[ 2 \left(E_{n} + { Z \over R} \right) \right]^{-3/2} 
= T \left[ 1 + { Z \over E_{n} R} \right]^{-3/2} 
\end{equation}
The orbital period is now a varying quantity function of time, 
and it is always $T' < T$.  
The exact value of the enhancement factor $T/T'$ depends upon $R$.
In \cite{pra} this enhancement factor was held constant, being the 
parameter $f_{T}$, usually taken equal to 2.
In order to have a quantitative estimate let us remark that captures 
occur preferentially for $R$ of order of $R'_{m}$ (see e.g. fig. 5 
of ref. \cite{pra}). We replace therefore $R$ with $R'_{m}$ in the 
previous equation and get that 
$T/T'$ reaches its minimum value 
$T/T' = (3/2)^{3/2} \approx 1.84$ for $Z = 1$ (with $\alpha = 1$).
The ratio increases rather slowly with $Z$: asimptotically it follows 
the scaling $T/T' \approx Z^{3/4}, Z \to \infty$; however,
it is already $T/T' > 2$ for all integer values $Z > 1$.
Therefore, we expect to have enhanced cross sections with respect to 
model I when dealing with highly charged projectile ions, while they 
should be-very slightly-depressed in collisions with singly charged ions.
This is a confirmation of the guess done in \cite{pra}, according to 
which $f_{T}$ was likely to be an increasing function of $Z$.

Equation (\ref{eq:prob}) must therefore be rewritten (without the 
factor $f_T$):
\begin{equation}
\label{eq:probprime}
P_l  = 1 - \exp \left( - \int_{- t_m}^{+ t_m} {N_{\Omega} \over T'} dt \right) 
\quad .
\end{equation}
Unfortunately, the integral in (\ref{eq:probprime}) can no longer be 
computed analytically; however, $\sigma$ is still easily numerically computed 
with only a few lines of code written in any mathematical software package.

\begin{figure}
\epsfxsize=12cm
\epsfbox{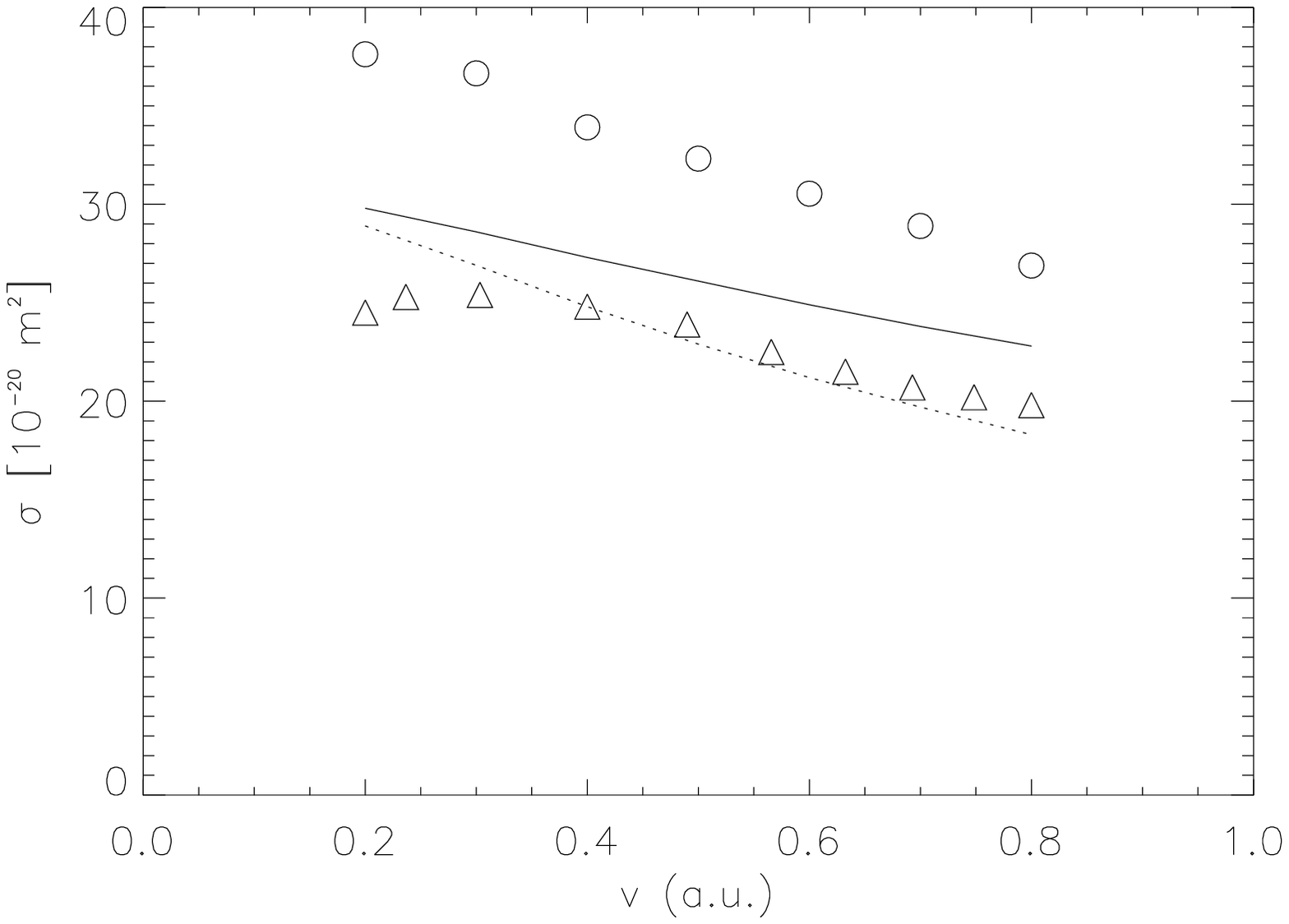}
\caption{Charge exchange cross section versus velocity for Be${}^{4+}$-H(1s) collisions. 
Triangles, data from ref. 6; circles, data from ref. 7; solid line, present 
model using $\alpha = 1$; dotted line, model I using $\alpha = 1, f_{T} = 2$}
\label{fig:hbe}
\end{figure}

We want now to test the model: as a first test case we address the process 
\begin{equation}
\label{eq:hbe}
{\rm H} + {\rm Be}^{4+} \to {\rm H}^{+} + {\rm Be}^{3+}  \quad . 
\end{equation} 
It has been studied by two different approaches in 
\cite{nf,adndt}, so we can rate predictions of Eqns. 
(\ref{eq:prob},\ref{eq:probprime}) against some sophisticated theories.
The results are plotted in fig. (\ref{fig:hbe}). The 
agreement between the old and the new model is rather good, with the 
latter slightly overestimating the former, as expected. 

\begin{figure}
\epsfxsize=12cm
\epsfbox{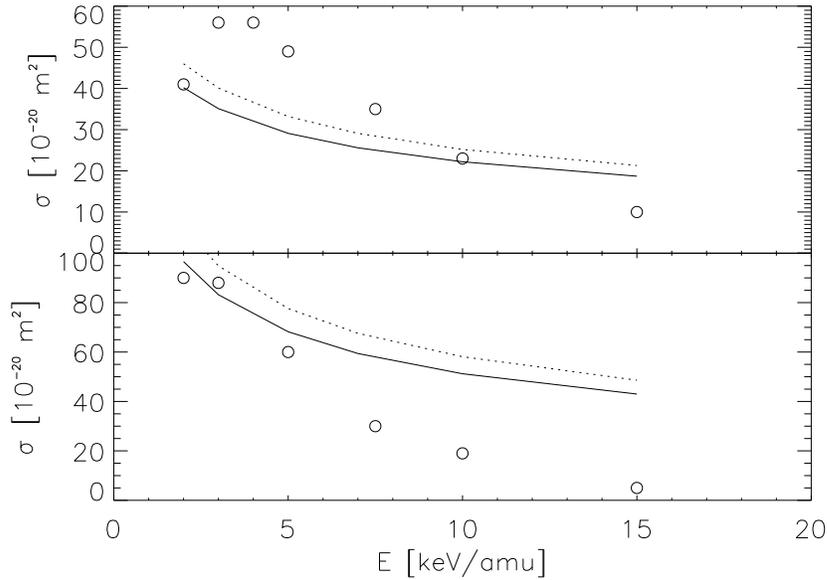}
\caption{Cross section for charge exchange in H${}^{+}$--Na(3s)
(upper) and  H${}^{+}$--Na(3p) (lower) collisions. 
Symbols, experimental data from ref. 8; solid line, present model; 
dotted line, model I.}
\label{fig:na3}
\end{figure}

As a second test case we present the results for collisions 
H${}^{+}$-- Na(3s,3p) (fig. \ref{fig:na3}).
Here the projectile is singly charged, so Eq. (\ref{eq:probprime})
is expected to give a result lower than Eq. (\ref{eq:prob}), and this is 
exactly found. 
In this case, as already remarked in \cite{pra}, the performance of 
the model is rather bad. We can just state again that the reason could 
be found in the non-hydrogen-like nature of the target. An upgrade of 
the model taking into account more realistic model potentials binding 
the electron could give remarkable enhancements. 

\begin{figure}
\epsfxsize=12cm
\epsfbox{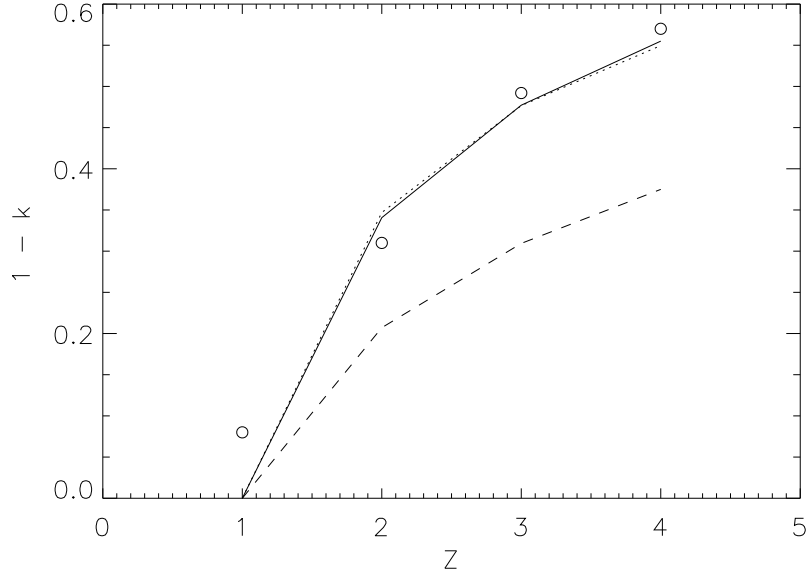}
\caption{Normalized energy defect as a function of projectile charge. 
Symbols, data from ref. 9; dashed line, OBM prediction from Eq. 
(17); dotted line, least squared fit to data using Eq. (16) and 
$\alpha = 1$; solid line, least squares fit to data using Eq. (16) 
and $\alpha = 2$.}
\label{fig:kappa}
\end{figure}

\begin{figure}
\epsfxsize=12cm
\epsfbox{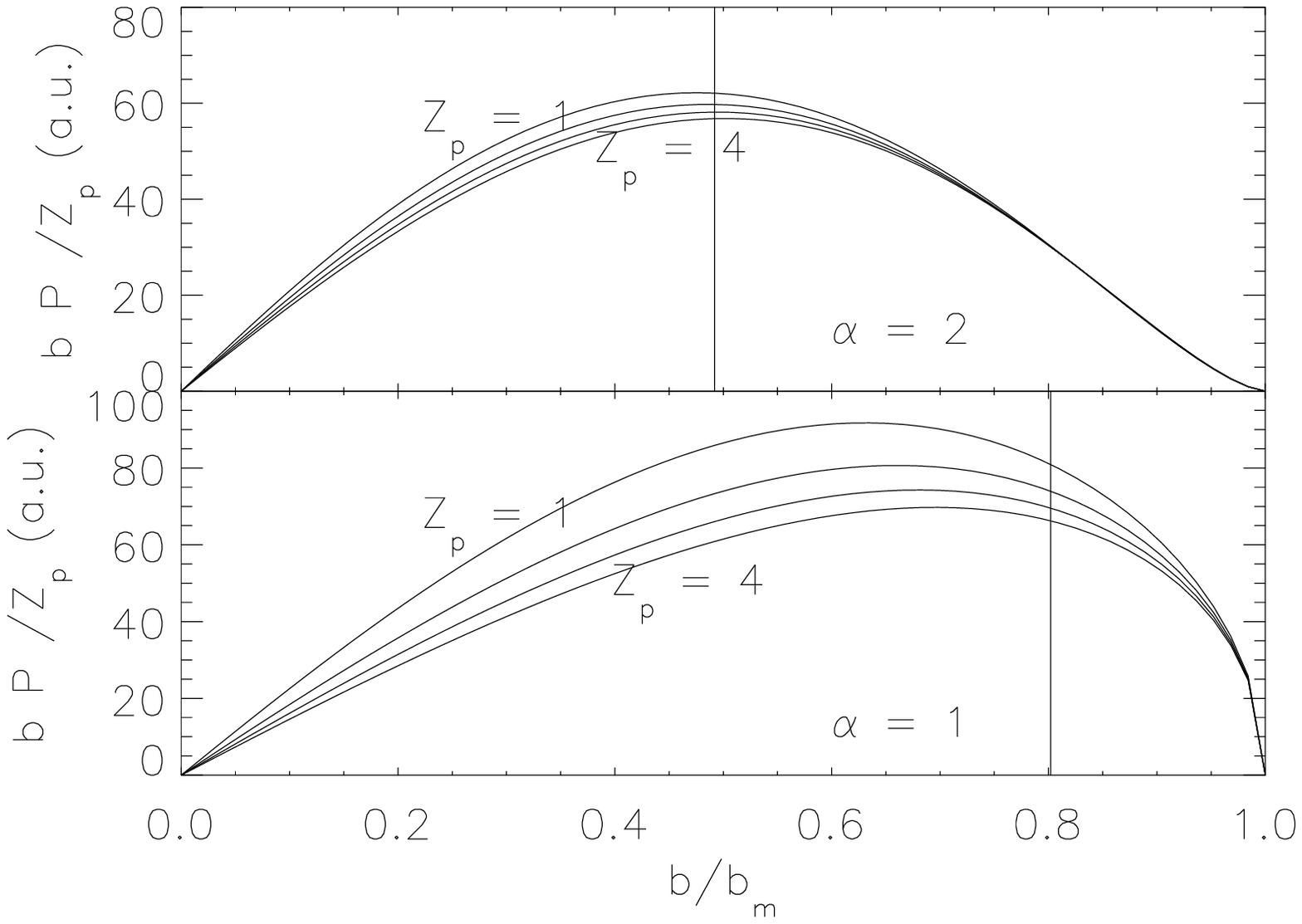}
\caption{Scaled differential cross section $ b P(b)/Z_{p}$ {\it versus} 
scaled impact parameter for the choices $\alpha = 2, f_{T} = 1$ (upper) and 
$\alpha = 1, f_{T} = 2$ (lower) and different projectile charges. 
The position of the maxima of the cross section 
as estimated by the least squares fit done using Eq. (16) are shown.}
\label{fig:pb}
\end{figure}

We want now to address a rather different point. It is 
partially unrelated with previous topics since it does not deal with 
any kind of improvement to the model. Instead, we will show that the 
OBM (any version of it, be the original version by Ostrovsky, the 
version I or the present one) is able to predict some experimental 
results previously thought not amenable to this kind of analysis. 
The experiments we are referring to, on charge exchange between slow ions and Rydberg 
atoms, are reported in the paper \cite{fisher}. 
Among other quantities, it was measured 
the binding energy of the captured electron $E_{p}$ as a function of the impact velocity $v$, 
of the projectile charge $Z_{p}$ and--above all--of the binding energy of the Rydberg 
target $E_t$, which allowed to compute the normalized energy defect function 
$1 - k = (E_{p} - E_{t})/E_{p}$. This is a convenient quantity since 
it can be computed for a number of models, including the
CTMC method and OBMs. Within the OBM the computation goes as follows:
the initial energy of the electron is $ E = - E_{t} - Z/R$ whereas in 
the final state it is $ E = - E_{p} - 1/R$. When the electron is being
transferred from one nucleus to the other the two quantities must be 
equal, thus
\begin{equation}
\begin{split}
\label{eq:kappa}
&E_{t} + { Z \over R} = E_{p} + {1 \over R} \to \\
&1 - k \equiv {E_{p} - E_{t} \over E_{p}} = { Z - 1 \over Z - 1 + E_{t} R} 
\quad .
\end{split}
\end{equation}
The maximal contribution to charge exchange is given by $R$ close to 
the maximum allowed $R'_{m}$ (see e.g. fig. 4 of \cite{ostrovsky} or 
fig. 5 of \cite{pra}). Therefore we set $ R  = f R'_{m}$. $f$ is a 
factor as yet undetermined accounting for the fact that the maximum is 
not exactly at $R'_{m}$ but at slightly lower values. Replacing this 
expression in (\ref{eq:kappa}) we get
\begin{equation}
\label{eq:fkappa}
1 - k = { Z - 1 \over Z - 1 + f (\alpha \sqrt{Z} + 1) } \quad .
\end{equation}
Naively, one could set $f = 1$ and get 
\begin{equation}
\label{eq:oldkappa}
 1 - k = { Z - 1 \over Z + 2 \sqrt{Z} } 
\end{equation}
(where we have also set $\alpha = 2$). This is the estimate for $1 - 
k $ as given in \cite{fisher} and also in \cite{ostrovsky,ryufuku}. 
The previous formula gives poor estimates for the experimental results 
and in \cite{fisher} it was suggested that the failure was due to the approximations
intrinsic to OBMs. We shall see, instead, that a little refinement to 
the above analysis gives us a rather good agreement with experiment. 
We exploit the extra degree of freedom given by $f$: 
a reasonable choice for $f$ is to choose the value of $R$ at which the capture cross 
section has a maximum and set $f = R/R'_{m}$. It is more convenient, 
although lesser accurate, to   
look for the maximum in $b P(b)$ as a function of $b$. 
Since the equation $d(b P)/db = 0$ cannot be solved analitically we  
resort to a backward procedure:  determine by 
a least squares fit the value of $f$ which best interpolates the data and 
check if this value corresponds to the maximum in $b P$.
In fig. (\ref{fig:kappa}) we plot the experimental data from ref. 
\cite{fisher}, the naive 
expression (\ref{eq:oldkappa}), and the above mentioned fits. 
Computations have been repeated for the two couples of parameter $\alpha = 
1, f_{T} = 2$ and $ \alpha = 2, f_{T} = 1$. 
For the computation of $P$ we have used expression 
(\ref{eq:prob}): using Eq. (\ref{eq:probprime}) would 
be a pointless complication. \\
Both fits are fairly good, although obtained with widely different values 
of $f$: the choice $\alpha = 1$ imposes $ f = 0.802$, while $\alpha = 
2$ yields $ f = 0.492$.
In fig. (\ref{fig:pb}) we plot the corresponding differential cross 
sections. The maximum of $ b P$ is only faintly a function of the 
projectile charge. The case $\alpha = 2$ gives a very good
accordance between the fit and the actually computed differential cross 
sections; thus, in this case, we can definitely state that the OBM is 
able to predict the results of \cite{fisher}. The case with $\alpha = 
1$ is slightly worse: the maximum of the cross section 
is around 0.65$\div$0.7.
   
To summarize, being able to justify one apparently free parameter from within the 
framework of the model itself is reassuring about its validity and 
its ability of catching as much physics of the capture process as 
possible. On the other hand, rather paradoxically, this makes even 
more puzzling the presence of the remaining free parameter, $\alpha$. 
We remind in fact that the choice $\alpha = 2$ should be the correct 
one, in that it is consistent with the same starting hypotheses
which allow us to arrive at Eq. (\ref{eq:periodprime}). It is however 
necessary using $\alpha = 1$ to be consistent with CTMC simulations, 
even though this means that we are making the same kind of error done 
when using $T$ instead of $T'$ \\
We have not at the moment a satisfactory explanation to this problem. 
It is not unlikely, however, that the ultimate reason lies in the 
failure of expression (\ref{eq:due}) for the electron energy $E$ close 
to the saddle point. That expression, in fact, holds rigorously only for large 
electron-projectile distances. At the saddle point, instead, the  
electron-target and electron-projectile distances are equivalent. 


\newpage

\begin{references}

\bibitem{niehaus} A. Niehaus, J. Phys. B: At. Mol. Phys. {\bf 19}, 2925 (1986). 

\bibitem{ostrovsky} V.N. Ostrovsky, J. Phys. B: At. Mol. Opt. Phys.
{\bf 28}, 3901 (1995).

\bibitem{jpb} F. Sattin, J. Phys. B: At. Mol. Opt. Phys. {\bf 33}, 
861, 2377 (2000).

\bibitem{pra} F. Sattin, Phys. Rev. A {\bf 62}, 042711 (2000).

\bibitem{landau} L.D. Landau and E.M. Lifshitz, {\it Quantum 
Mechanics} (Oxford, Pergamon, 1977).

\bibitem{nf} P.S. Krstic, M. Radmilovic and R.K. Janev, {\it Atomic 
and Plasma-Material Data for Fusion} (IAEA, Vienna, 1992), vol. 3, p. 
113.

\bibitem{adndt} C. Harel, H. Jouin and B. Pons, At. Data Nucl. Data Tables
{\bf 68}, 279 (1998).  

\bibitem{na3} J.W. Thomsen {\it et al}, Z. Phys. D {\bf 37}, 133 (1996).

\bibitem{fisher} D.S. Fisher {\it et al}, Phys. Rev. Lett. {\bf 81}, 
1817 (1998).

\bibitem{ryufuku} H. Ryufuku, K. Sasaki and T. Watanabe, Phys. Rev. 
A {\bf 21}, 745 (1980).


\end{references}
\end{document}